\documentclass[12pt]{article}
\usepackage{a4wide}
\usepackage{amssymb}
\begin{document}
{\renewcommand{\thefootnote}{\fnsymbol{footnote}}
%\hfill  IGC--yy/m--n\\
\medskip
\begin{center}
  {\LARGE Effective Casimir Conditions\\[3mm] and Group
    Coherent States }\\
  \vspace{1.5em} Martin Bojowald$^1$\footnote{e-mail address: {\tt
      bojowald@gravity.psu.edu}} and Artur Tsobanjan$^{1,2}$\footnote{e-mail
    address: {\tt artur.tsobanjan@gmail.com}} 
  \\
  \vspace{0.5em}
  $^1$ Institute for Gravitation and the Cosmos,
  The Pennsylvania State
  University,\\
  104 Davey Lab, University Park, PA 16802, USA\\
  \vspace{0.5em}
 $^2$ American University, 4400 Massachusetts Ave NW, Washington, DC 20016,
 USA\\ 
  \vspace{1.5em}
\end{center}
}

\setcounter{footnote}{0}

\newtheorem{theo}{Theorem}
\newtheorem{lemma}{Lemma}
\newtheorem{defi}{Definition}

\newcommand{\proofend}{\raisebox{1.3mm}{\fbox{\begin{minipage}[b][0cm][b]{0cm}
\end{minipage}}}}
\newenvironment{proof}{\noindent{\it Proof:} }{\mbox{}\hfill \proofend\\\mbox{}}
\newenvironment{ex}{\noindent{\it Example:} }{\medskip}
\newenvironment{rem}{\noindent{\it Remark:} }{\medskip}

\newcommand{\case}[2]{{\textstyle \frac{#1}{#2}}}
\newcommand{\lP}{\ell_{\mathrm P}}

\newcommand{\md}{{\mathrm{d}}}
\newcommand{\tr}{\mathop{\mathrm{tr}}}
\newcommand{\sgn}{\mathop{\mathrm{sgn}}}

\newcommand*{\R}{{\mathbb R}}
\newcommand*{\N}{{\mathbb N}}
\newcommand*{\Z}{{\mathbb Z}}
\newcommand*{\Q}{{\mathbb Q}}
\newcommand*{\C}{{\mathbb C}}

\begin{abstract}
  Properties of group coherent states can be derived ``effectively'' without
  knowing full wave functions. The procedure is detailed in this article as an
  example of general methods for effective constraints. The role of
  constraints in the present context is played by a Casimir condition that
  puts states within an irreducible representation of a Lie group (or,
  equivalently, on a quantization of a co-adjoint orbit of the dual Lie
  algebra). Simplifications implied by a Casimir condition, compared with
  general first-class constraints, allows one to show that the correct number
  of degrees of freedom is obtained after imposing the condition. When
  combined with conditions to saturate uncertainty relations, moments of group
  coherent states can be derived. A detailed example in quantum cosmology
  (cosmic forgetfulness) illustrates the usefulness of the methods.
\end{abstract}

\section{Introduction}

There are several different definitions of coherent states based on
group-theoretical properties, which all play important roles in diverse areas
of theoretical physics. Explicit examples of wave functions of coherent states
can show aspects of quantum evolution and semiclassical phenomena in a clean
way, but it is not always possible to find such explicit realizations in
general-enough terms. However, physical properties can still be extracted if
one works with moments of a state instead of wave functions, a description
which is the basis also of canonical effective theory
\cite{EffAc,Karpacz}. Dynamical equations for the moments follow from
generalizations of Ehrenfest's equations, while coherence is implemented by
the condition that moments saturate uncertainty relations. The main link to
group coherent states, studied in this article, is the imposition of an
additional Casimir condition that restricts moments to states that belong to
an irreducible representation of a Lie group describing the quantum system (or
the quantization of a co-adjoint orbit in the dual Lie algebra of the
group). 

Imposing the Casimir to be constant can be interpreted as a constraint on the
original (non-symplectic) phase space with a Poisson structure given by the
dual Lie algebra, suggesting that methods for effective constraints
\cite{EffCons,EffConsRel} can be used.  Because a Casimir operator $\hat{C}$
commutes with all other operators, additional simplifications compared with
general first-class constraints arise: (i) expectation values
$\langle\hat{O}\hat{C}\rangle$, which feature prominently in effective
constraints, equal the symmetric version
$\frac{1}{2}\langle\hat{O}\hat{C}+\hat{C}\hat{O}\rangle$ and are guaranteed to
be real for self-adjoint $\hat{O}$ and $\hat{C}$, and (ii) no gauge flow
$\delta\langle\hat{O}\rangle/\delta \epsilon=
\langle[\hat{O},\hat{C}]\rangle/i\hbar$ generated by a Casimir constraint need
be considered. (On a non-symplectic phase space, first-class constraints do
not necessarily generate gauge flows.) These simplifications allow us to
confirm, to all orders in a semiclassical or moment expansion, that the
correct number of quantum degrees of freedom is left after imposing the
Casimir condition.

In a second step, we then combine the Casimir condition with the requirement
that uncertainty relations be saturated, restricting moments to those of a
group coherent state. Higher orders of moments are more difficult to manage at
this level, but we will be able to demonstrate several interesting
relationships between the different conditions imposed.

Quantum cosmology presents an example in which standard group coherent states
are not always available in general terms, that is with full squeezing, while
the effective methods elaborated here do apply. We end our article with a
detailed discussion, clarifying the contentious issue of cosmic forgetfulness
\cite{BeforeBB} which posits that certain pre-big bang models in loop quantum
cosmology \cite{LivRev} suffer from a severe lack of control on the pre-big
bang state. This issue has lost its urgency with the recent discovery that the
same models, when embedded consistently in a setting with perturbative
inhomogeneity, lead to signature change at high density
\cite{Action,ScalarHol}, so that no state can be evolved deterministically
through the big bang. Nevertheless, the issue of cosmic forgetfulness may
still be of interest from a mathematical perpective.

\section{Effective Casimir constraints}
\label{s:EffCas}

For simplicity, we consider the case of a single Casimir condition $C$,
imposed on a non-symplectic phase space so that the submanifold $C={\rm
  const}$ is symplectic. The phase-space function $C$ can be seen as one
coordinate of a Poisson manifold, such that $C$ itself is a Casimir function
in the Poisson sense, and submanifolds $C={\rm const}$ are the symplectic
leaves. Since $C$, by definition, has a vanishing Poisson bracket with any
other function on the Poisson manifold, it does not generate a Hamiltonian
gauge flow when viewed as a constraint. Nevertheless, it can be identified as
a first-class constraint owing to the non-symplectic nature of the phase
space. (Notions equivalent for symplectic geometry and usually associated with
first-class constraints, such as the properties of non-trivial gauge flows and
symplectic properties of constraint sets, may no longer be equivalent for
Poisson manifolds.  The generalization of standard definitions therefore
requires some care \cite{brackets}.)

As an example, we may look at a 3-dimensional manifold equipped with
coordinates (or basic functions) $V,J_+,J_-$ and Poisson brackets
$\{V,J_+\}=J_-$, $\{V,J_-\}= -J_+$ and $\{J_+,J_-\}= -V$, for instance
interpreted as the (dual) Lie algebra of ${\rm sl}(2,{\mathbb R})$. The
Casimir function reads $C=J_+^2+J_-^2-V^2$. Another way to interpret the same
system is to use partially complex variables $V$ with $J=J_++iJ_-$ and $J^*=
J_+-iJ_-$. One can then realize the Poisson brackets
\begin{equation}\label{PBComplex}
 \{V,J\}=-iJ\quad,\quad \{V,J^*\}=iJ^*\quad,\quad \{J,J^*\}=2iV
\end{equation}
by functions $V$ and $J:=V\exp(-iP)$ of
canonical variables $V$ and $P$ with $\{V,P\}=1$.  Setting the Casimir
function to zero, $C=JJ^*-V^2=0$, then amounts to a reality condition for
$P$. In this form, the Casimir condition plays a role in some cosmological
models \cite{BouncePert,BounceCohStates}.

Upon passing on to the quantum treatment of the system, the role of $C$ as a first-class constraint (though one on a
non-symplectic phase space) allows us to employ the effective-constraint
methods developed in \cite{EffCons,EffConsRel,EffConsComp}. To this end, we
view the corresponding quantum system algebraically, based on the commutators
of basic operators quantizing (\ref{PBComplex}). The state space of the
algebra can be formulated geometrically by making use of the expectation-value
functional $\langle\cdot\rangle$, applied to all polynomials in basic
operators. Expectation values of products of operators are identified as
moments
\begin{equation} \label{Moments}
 \Delta(V^aJ_+^bJ_-^c):=\langle (\hat{V}-\langle\hat{V}\rangle)^a
 (\hat{J}_+-\langle\hat{J}_+\rangle)^b
 (\hat{J}_--\langle\hat{J}_-\rangle)^c\rangle_{\rm
   Weyl-ordered}
\end{equation}
of the basic operators in a state considered. Any product that is not
Weyl-ordered can be rearranged as a sum of Weyl-ordered terms, some of which
with explicit $\hbar$-factors. The set (\ref{Moments}) therefore
prescribes the expectation values for all polynomials in the basic variables. A Poisson geometry of the quantum phase space, given by expectation
values and moments of basic operators, is provided by the commutator, defining
a Poisson bracket on expectation-value functionals by
\begin{equation} \label{Poisson}
 \{\langle\hat{A}\rangle,\langle\hat{B}\rangle\}=
\frac{\langle[\hat{A},\hat{B}]\rangle}{i\hbar}\,,
\end{equation}
extended to all functions by the Leibniz rule.

One element of this algebra is the Casimir or constraint operator $\hat{C}$ to
be imposed as a constraint, $\hat{C}|\psi\rangle=0$. (For a non-zero Casimir,
we can simply redefine the constraint operator as $\hat{C}-{\rm
  const}\hat{\mathbf{1}}$ without changing commutators.) Evaluated in any state annihilated by $\hat{C}$, the expectation values $\langle\widehat{\rm
  pol}\hat{C}\rangle$ must all vanish for arbitrary polynomials
$\widehat{\rm pol}$ in the basic operators. For one quantum constraint
operator we obtain an infinite number of constraint functions on the quantum
phase space. For multiple ones, if the quantum constraints are first class,
the system of effective constraints is first-class in the phase-space sense
\cite{EffCons}. In the special case of a Casimir constraint, all effective
constraints are Casimir functions on the Poisson manifold defined by
(\ref{Poisson}): it is straightforward to see that a constraint of the form
$\langle\widehat{\rm pol}\hat{C}\rangle=0$ weakly Poisson commutes with all
quantum phase-space functions if $\hat{C}$ commutes with all operators.

For first-class constraints on symplectic phase spaces, the viability of these
methods has already been demonstrated, addressing also the problem of time
\cite{EffTime,EffTimeLong,EffTimeCosmo}. In general, the ordering of effective
constraints in the specific form 
\begin{equation}\label{Cpol}
 C_{\rm pol}:=\langle\widehat{\rm pol}\hat{C}\rangle
\end{equation}
is important for the system of constraints in order to remain first class and to vanish in
 physical states. We may assume symmetric polynomials without loss of
generality, because re-ordering terms would just contribute quantities
proportional to lower-order constraints. However, even if the basic operators and
$\hat{C}$ are assumed to be self-adjoint with respect to some $*$-relation on
the basic algebra, which we will do in what follows, the ordering in
(\ref{Cpol}) is in general not symmetric, leading to the possibility of
complex-valued effective constraints. For Casimir conditions, on the other
hand, we have $[\widehat{\rm pol},\hat{C}]=0$ by definition, so that we can
substitute the symmetric ordering $\frac{1}{2} \langle \widehat{\rm
  pol}\hat{C}+ \hat{C}\widehat{\rm pol}\rangle$ without changing the
expectation value. We are therefore dealing with real-valued effective
constraints for all polynomial or moment orders. Moreover, Casimir conditions
are easier to implement because no gauge flows need be considered: we have a
vanishing
\[
 \{\langle\hat{O}\rangle,\langle\widehat{\rm
  pol}\hat{C}\rangle\} = \frac{\langle [\hat{O},\widehat{\rm
    pol}]\hat{C}\rangle}{i\hbar}\approx 0
\]
on the solution space of the effective constraints, again using the
commutation property of a Casimir.  This simplification is the main reason
that allows us to directly test effective methods at higher orders in moments.

\subsection{Removing degrees of freedom}

In classical systems, Casimir constraints remove phase-space degrees of freedom so that, in the
absence of a gauge flow, the constraint surface is symplectic. The dimension
of a symplectic manifold is restricted to be even, and its quantization
corresponds to a fixed pattern of expectation values and moments: for every
canonical pair $(q,p)$ there are two independent expectation values and,
starting with $n=2$ reaching ad infinitum, a tower of $n+1$ moments for every
integer $n$, defined as in (\ref{Moments}) for a single canonical pair of
basic operators:
\begin{equation} \label{MomentsCan}
 \Delta(q^ap^b):=\langle (\hat{q}-\langle\hat{q}\rangle)^a
 (\hat{p}-\langle\hat{p}\rangle)^b\rangle_{\rm
   Weyl-ordered}\,.
\end{equation}
Semiclassically, a moment of order $n$ behaves like $O(\hbar^{n/2})$, giving
rise to a general semiclassical expansion with a finite number of degrees of
freedom per canonical pair at any fixed order $n$. If this pattern is
violated, one cannot interpret the degrees of freedom in the usual way of
quantum mechanics, indicating either extra constraints if there are not enough
free moments or spurious degrees of freedom if too many variables remain
unrestricted. In particular, for a single Casimir constraint we must, by
imposing $C_{\rm pol}=0$, eliminate sufficiently many variables to leave a
certain number of canonical pairs with their characteristic moments.

For the example of one quadratic Casimir constraint on a 3-dimensional Poisson
manifold such as the one of $(V,J_+,J_-)$, one can heuristically see that the
reduction is correct, working locally and assuming that all effective
constraints are independent. Taking an expectation value
$\langle\hat{C}\rangle=0$ restricts the basic expectation values to two
independent ones (for fixed second-order moments). The next order of
constraints is obtained for $\langle\hat{A}\hat{C}\rangle=0$ with $\hat{A}$
one of the three basic operators (or a linear combination of them). Fixing
third-order moments, we obtain three constraints for the six second-order
moments of three independent basic operators, just enough to restrict the
second-order moments to three independent ones: two fluctuations and one
correlation. To see the correct reduction in numbers for all orders, we note
that one can, at least locally, view the Casimir function as a coordinate on
the Poisson manifold transversal to symplectic leaves. Instead of using basic
phase-space functions such as $V$, $J_+$ and $J_-$, we can locally transform
to Casimir--Darboux coordinates of a canonical pair $(q,p)$ on the symplectic
leaf and $C$ transversally. This decomposition is relevant also for the
counting of degrees of freedom of the corresponding quantum system. As for
constraints, in addition to the expectation value of $\hat{C}$, we must remove
all moments of $\hat{C}$ with itself (such as the fluctuation $\Delta C$) as
well as all cross moments with $\hat{q}$ and $\hat{p}$. These cross moments
are nothing but the effective constraints $\langle\widehat{\rm
  pol}\hat{C}\rangle=0$, with $\widehat{\rm pol}$ a polynomial of degree $n$
to restrict $C$-moments of order $n+1$. Since there are as many cross-moments
as constraints of this form, we obtain the correct number of degrees of
freedom provided the constraints are all independent (and the local argument
remains valid in quantum theory).

In most cases, and always if the coordinate change to Casimir--Darboux
coordinates is not global, the transformation from $(V,J_+,J_-)$ to $(q,p,C)$
is non-linear and relationships between $(V,J_+,J_-)$-moments and $(q,p,C)$
are non-trivial. For instance, a moment of low order in one system may involve
moments of all orders in the other. For a global statement about parameter
counting, more-detailed considerations must be performed. 

\subsection{Counting truncated constraint conditions}

For more generality, we work with a general finite-dimensional algebra of
generators $\hat{x}_i$, $i=1, 2, \ldots M$, denoting their expectation values
(or, occasionally, the corresponding classical values) by $x_i$, $i=1, 2,
\ldots M$. To define the moments, we introduce
\[
\widehat{\Delta x}_i := \hat{x}_i - x_i \hat{\mathbf{1}}
\]
and their Weyl-ordered products
\[
\hat{e}_{\vec{i}} = \hat{e}_{(i_1, i_2, \ldots i_M)} :=
\left(\widehat{\Delta x}_1^{i_1} \widehat{\Delta x}_2^{i_2} \ldots
\widehat{\Delta x}_M^{i_M} \right)_{\rm Weyl-ordered}\,,
\]
which form a linear basis for the (extended) algebra.  We use a compact
notation in which $\vec{i}$ is an $M$-tuple of non-negative integers.
Expectation values
\begin{equation}\label{Momentsx}
\Delta(\vec{x}^{\vec{i}}):=\Delta(x_1^{i_1}\cdots x_M^{i_M}) =
%\varepsilon_{(i_1, i_2, \ldots i_M)} =  \varepsilon_{\vec{i}} =
\langle \hat{e}_{\vec{i}} \rangle\,.
\end{equation}
of the basis elements are the moments. For later use we define the degree
$| \vec{i} | := \sum_{n=1}^M i_n$ and a partial ordering $\vec{i} \geq
\vec{j}$ if $i_n \geq j_n$ for all $n$, so that $\vec{i} > \vec{j}$ if
$\vec{i} \geq \vec{j}$ and $\vec{i} \neq \vec{j}$.  We will use $\vec{i}!$ to
denote $(i_1!i_2!  \ldots i_M!)$ and, as already defined in (\ref{Momentsx}),
$\vec{x}^{\vec{i}}=x_1^{i_1}\cdots x_M^{i_M}$. By construction,
%$\varepsilon_{\vec{i}}=0$, 
$\Delta(\vec{x}^{\vec{i}})=0$ for all $\vec{i}$ with $| \vec{i}| = 1$.

Interpreting the $\hat{x}_i$ as basic operators of a quantum system, we assume
that their commutator algebra is linear (and follows from a direct quantization
of Poisson brackets of the corresponding classical functions):
\begin{equation}\label{Algebra}
 [\hat{x}_i, \hat{x}_j] = i\hbar \epsilon_{ij}^{\ \ k} \hat{x}_k
\end{equation}
where $\epsilon_{ij}^{\ \ k}$ are structure constants, identical to the
structure constants of the classical Poisson algebra. For a semisimple Lie
algebra, which we will assume in an example later in this article,
$\epsilon_{ijk}$ with the third index pulled down by contraction with the
Killing metric, are totally antisymmetric. The case of the Weyl algebra
$[\hat{q}, \hat{p}]=i\hbar \hat{\mathbf{1}}$ for a single canonical pair, as
another important example, does not seem to fall within the current
setting. However, we could treat $\hat{q}$, $\hat{p}$ and $\hat{\mathbf{1}}$
as three `generators' or, alternatively, start from the Heisenberg algebra
$[\hat{q},\hat{p} ]= i\hbar \hat{z}$ and enforce a constraint $\hat{C} =
\hat{z} - \hat{\mathbf{1}}$ in the way described below. 

It follows that
\begin{equation} \label{DeltaComm}
\left[ \widehat{\Delta x_i}, \widehat{\Delta x_j} \right] = i\hbar
\sum_k \epsilon_{ij}^{\ \ k} \hat{x}_k = i \hbar \sum_k
\epsilon_{ij}^{\ \ k} \left( \widehat{\Delta x_k} + x_k
\hat{\mathbf{1}} \right)
\end{equation}
with the two right-hand-side terms respectively of polynomial degree $1$ and $0$ in the operators $\widehat{\Delta x_i}$. (This relation
is not formulated for linear operators, owing to the presence of expectation
values that depend on a state. Nevertheless, the usual rules can be applied if
one treats $x_k$, during a calculation, as some real number and identifies it
with the expectation value of $\hat{x}_k$ only in the final results.) In this
form, the relation plays an important role in considerations of orderings: it
allows one to apply commutators to symmetrically order any product of
$\widehat{\Delta x}_i$-s by adding terms of lower polynomial degree and
proportional to powers of $\hbar$,
\begin{eqnarray*}
\hat{e}_{\vec{i}} \hat{e}_{\vec{j}} &=& \hat{e}_{(\vec{i}+\vec{j})}
+ \hbar \sum_{\tiny{\begin{array}{c} \vec{k} < \vec{i}+\vec{j} \\
|\vec{k}| \leq |\vec{i}+\vec{j}|-1 \end{array}}} {}^{(1)}\!\beta_{\vec{i},
\vec{j}}{}^{\vec{k}} \:\hat{e}_{\vec{k}} + \hbar^2 \sum_{\tiny{\begin{array}{c} \vec{k} < \vec{i}+\vec{j} \\
|\vec{k}| \leq |\vec{i}+\vec{j}|-2 \end{array}}} {}^{(2)}\!\beta_{\vec{i},
\vec{j}}{}^{\vec{k}}\:\hat{e}_{\vec{k}} + \cdots
\end{eqnarray*}
where ${}^{(n)}\!\beta_{\vec{i},
\vec{j}}{}^{\vec{k}}$ are polynomials in
the expectation values $x_i$.

\subsubsection{Constraints and truncation} \label{sec:truncation}

Following the general procedure of effective constraints, the Casimir
condition of an operator $\hat{C}$ is imposed by demanding $\langle
f(\hat{x}_1, \hat{x}_2, \ldots, \hat{x}_M) \hat{C} \rangle = 0$ for all
polynomial functions $f$. These are infinitely many conditions for infinitely
many moments, but working order by order in the moments (or in a semiclassical
expansion) one can truncate these conditions systematically using the basis
$\{\hat{e}_{\vec{i}} \}$. We introduce
\[
C_{\vec{i}} := \langle \hat{e}_{\vec{i}\ } \hat{C} \rangle = 0, \ \
\forall \ \vec{i} \in \mathbb{Z}_+^M
\]
with the convention $\hat{e}_0=\hat{\mathbf{1}}$, so that $C_0 := \langle
\hat{C} \rangle$. In order for the truncation to be consistent, we must
suitably combine the truncation of constraint functions according to the
degree $|\vec{i}|$ with a truncation of variables that feature in the system.

Assume that we truncate at some order $N \geq 2$ of the semiclassical
expansion: we drop moments of degree greater than $N$, that is all
$\Delta(\vec{x}^{\vec{i}})$
%$\varepsilon_{\vec{i}}$ 
with $| \vec{i} | > N$. The truncation of the system
of constraints is more subtle. All $C_{\vec{i}}$ are linear functions of the
moments 
$\Delta(\vec{x}^{\vec{j}})$,
%$\varepsilon_{\vec{j}}$, 
but they contain terms of three different
types. It is useful to assign orders to these terms according to their type,
as follows:
\begin{itemize}
\item A term of the form 
$f(x_1, x_2 \ldots x_M)\Delta(\vec{x}^{\vec{i}})$,
%$f(x_1, x_2 \ldots x_M)\varepsilon_{\vec{i}}$, 
where
  $f$ is a polynomial in the expectation values not proportional to the classical
  constraint, is assigned semiclassical order equal to $| \vec{i} |$.
\item A term of the form 
$C(x_1, x_2 \ldots x_M)\Delta(\vec{x}^{\vec{i}})$,
%$C(x_1, x_2 \ldots x_M)\varepsilon_{\vec{i}}$, 
where
  $C$ is the classical polynomial expression for the constraint, requires a
  special treatment. As an exception to the previous point, it is assigned
  semiclassical order $| \vec{i} |+ 2$. This exception can be understood from
  the fact that $C(x_1, x_2 \ldots x_M)$, although it may not vanish exactly
  when quantum corrections are included in $C_0=C+O(\hbar)$, is of the order
  $\hbar$ when the quantum constraint $C_0=0$ is imposed. See also
  \cite{EffCons}.
\item Terms such as 
$\hbar^n f(x_1, x_2 \ldots x_M)\Delta(\vec{x}^{\vec{i}})$,
%$\hbar^n f(x_1, x_2 \ldots x_M)\varepsilon_{\vec{i}}$,
  which may arise upon reordering algebra elements, are assigned
  semiclassical order equal to $| \vec{i} |+ 2n$.
\end{itemize}
All these orders are consistent with the correspondence of order $n$ to terms
$O(\hbar^{n/2})$ as it applies to the moments.  To truncate, terms in any
expression for a constraint function are dropped when they are of
semiclassical order higher than $N$.

\subsubsection{Counting degrees of freedom}

We begin by counting the degrees of freedom of an unconstrained system
generated by polynomials in $M$ basic variables (for a phase space of
dimension $M$). We have $M$ expectation values and, at each semiclassical
order $N \geq 2$, the degrees of freedom are represented by the independent
functions 
$\Delta(\vec{x}^{\vec{i}})$
%$\varepsilon_{\vec{i}}$ 
with $| \vec{i} | = N$. 

The task of counting degrees of freedom is simplified by rephrasing the
problem: The number of such variables is given by the number of $M$-tuples of
non-negative integers with $|\vec{i}|=N$, a quantity that we will call
$\mathcal{N}_M(N)$. Each such $M$-tuple is produced by considering a row of
$N+M-1$ identical objects, marking $M-1$ of them to serve as partitions. The
value of $i_n$ is then the number of unmarked objects between partition
$(n-1)$ and partition $n$. (Partitions $0$ and $M$ are assumed to be at the
ends.) With this rephrasing we directly obtain
\[
\mathcal{N}_M(N) = {N+M-1 \choose M-1}\,.
\]

Each constraint, in the absence of gauge flows, removes a single classical
degree of freedom, and we expect the tower of effective constraint conditions
to remove the corresponding moments.  After the constraints are imposed, there
should be as many degrees of freedom as for a system with $M-1$ generators,
with $\mathcal{N}_{M-1}(N) = {N+M-2 \choose M-2}$ free variables at each
semiclassical order. Using the identity ${a\choose b} - {a-1 \choose b} =
{a-1\choose b-1}$, we conclude that the required number of independent
conditions at each order should be ${N+M-2 \choose M-1}$.

In order to show consistency of the effective procedure, we now proceed to
counting the number of conditions independently.  This question is more
difficult to answer because constraint conditions $C_{\vec{i}}$ generally mix
terms of different orders. As a first step, we show that, truncated at a given
order, the system of constraints is finite. The number of constraints at
each order is then the number of additional non-trivial constraint conditions
that arise when we raise the truncation order by one. 

A general constraint function has the form
\begin{eqnarray}\label{eq:constraints}
C_{\vec{i}} &=& \sum_{\vec{j} \geq \vec{i}} \frac{1}{(\vec{j} -
\vec{i})!} \frac{\partial^{|\vec{j} - \vec{i}|} C}{\partial
x_1^{j_1-i_1} \ldots \partial x_{M}^{j_M-i_M}} \Delta(\vec{x}^{\vec{j}})
+ \hbar \sum_{|\vec{j}| \geq |\vec{i}|-1} {}^{(1)}\!\alpha
_{C_{\vec{i}}}{}^{\vec{j}}\: \Delta(\vec{x}^{\vec{j}}) \nonumber \\ & & + \hbar^2
\sum_{|\vec{j}| \geq |\vec{i}|-2} {}^{(2)}\!\alpha
_{C_{\vec{i}}}{}^{\vec{j}} \:\Delta(\vec{x}^{\vec{j}}) + \ldots + \hbar^{|\vec{i}|}
\sum_{\vec{j}} {}^{(|\vec{i}|)}\!\alpha
_{C_{\vec{i}}}{}^{\vec{j}}\:
\Delta(\vec{x}^{\vec{j}})
\end{eqnarray}
%\begin{eqnarray}\label{eq:constraints}
%C_{\vec{i}} &=& \sum_{\vec{j} \geq \vec{i}} \frac{1}{(\vec{j} -
%\vec{i})!} \frac{\partial^{|\vec{j} - \vec{i}|} C}{\partial
%x_1^{j_1-i_1} \ldots \partial x_{M}^{j_M-i_M}} \varepsilon_{\vec{j}}
%+ \hbar \sum_{|\vec{j}| \geq |\vec{i}|-1} \alpha_{C_{\vec{i}}}^{(1)
%\ \vec{j}} \varepsilon_{\vec{j}} \nonumber \\ & & + \hbar^2
%\sum_{|\vec{j}| \geq |\vec{i}|-2} \alpha_{C_{\vec{i}}}^{(2) \
%\vec{j}} \varepsilon_{\vec{j}} + \ldots + \hbar^{|\vec{i}|}
%\sum_{\vec{j}} \alpha_{C_{\vec{i}}}^{(|\vec{i}|) \ \vec{j}}
%\varepsilon_{\vec{j}}
%\end{eqnarray}
Here ${}^{(n)}\!\alpha
_{C_{\vec{i}}}{}^{\vec{j}}$ are coefficients of
semiclassical order zero, polynomial in the expectation values $x_i$. The first
sum comes from the Weyl-symmetric part of the element $\hat{e}_{\vec{i}\ }
\hat{C}$. Subsequent sums arise from its components that are antisymmetric in
one, two and more adjacent pairs of moment-generating elements
$\widehat{\Delta x}_i$. Each antisymmetric pair can be reduced by using the
commutation relations, producing the powers of $\hbar$. For a Casimir
constraint, $\hat{C}_{\vec{i}}$ is guaranteed to be real; therefore, there
must be an even number of commutators applied in each re-ordering step and we
have only even powers of $\hbar$, or ${}^{(n)}\!\alpha
_{C_{\vec{i}}}{}^{\vec{j}}=0$ for odd $n$.

The important feature of the above expansion is that the lowest semiclassical
order terms are 
$C\Delta(\vec{x}^{\vec{i}})$
%$C\varepsilon_{\vec{i}}$
and 
\begin{equation} \label{Cbar}
\bar{C}_{\vec{i}}:= \sum_k \frac{\partial
  C}{\partial x_k}
\Delta(\vec{x}^{(i_1,\ldots,i_{k-1},i_k+1,i_{k+1},\ldots,i_M)})\,.
\end{equation}
%\varepsilon_{(i_1, \ldots, i_k+1, \ldots, i_M)}$. 
The
latter term has the lowest order, $|\vec{i}|+1$, keeping in mind that for the
purposes of truncating the constraints, $C$ is of order $2$. After truncation
at order $N$, constraints $C_{\vec{i}}=0$ are satisfied identically for all
$|\vec{i}| >N-1$. This observation allows us again to rephrase the counting
problem: The number of non-trivial conditions up to order $N$ is the same as
the number of non-negative integer $M$-tuples of degree $N-1$ and less.  As we
go from truncation at order $N-1$ to truncation at order $N$, this number
changes by the number of $M$-tuples of degree $N-1$, which is exactly the
quantity $\mathcal{N}_M(N-1)={N+M-2 \choose M-1}$ required to be eliminated by
the counting of degrees of freedom.

Thus, provided the non-trivial constraint conditions remaining after
truncation are functionally independent, they remove precisely one
combinatorial degree of freedom in the quantum mechanical sense.  In order to
show independence, which is done in detail in \cite{Counting}, one considers
the gradients $\{{\rm d}_{\Delta}\bar{C}_{\vec{i}} \}_{1\leq |\vec{i}|\leq
  N-1}$ on the space of expectation values and moments. While the operator
$d_{\Delta}$\ takes the gradient with respect to the quantum moments only, the
functional independence conditions derived depend predominantly on the
classical form of the constraint function. For a semiclassical state, applying
careful truncations as before, one can conclude: (i) The truncated constraints
are independent as long as ${\rm d}C$ (the gradient taken with respect to the
expectation values only) is not comparable to $\hbar$ or the moments in at
least one coordinate direction, assuming that expectation values satisfy the
classical constraint. This condition can be interpreted as the semiclassical
analog of regularity of the constraints. (ii) For expectation values off the
classical constraint surface, the constraint functions are functionally
independent as long as, for some $k$, neither $\partial C/\partial x_k$ nor
$\partial^{N-2}(C^{-1})/\partial x_k^{N-2}$ are comparable to $\hbar$ or the
moments.

The conditions may clearly be violated somewhere on the quantum phase
space, but near the classical constraint surface terms such as $1/C$
and its derivatives diverge and appear in the gradients. The
gradients, unlike the constraints, can thus be considered ``large.''
These conditions are sufficient, but not necessary, so that in some
cases the constraints may be independent even if the conditions do not
hold. To summarize, so long as the classical constraint is sufficiently regular, the truncated set of gradients $\{ {\rm d}_{\Delta}\bar{C}_{\vec{i}}
\}_{1\leq |\vec{i}|\leq N-1}$ is linearly independent for expectation
values $x_i$ lying in some neighborhood of the classical constraint
surface $C=0$, leading to functional independence of the truncated set of constraint functions in that region.

\section{Uncertainty relations}

For moments of a group coherent state, in a quantization of a co-adjoint orbit
on the dual of the Lie algebra corresponding to (\ref{Algebra}), we require
that uncertainty relations be saturated. Together with the Casimir condition,
several equations are then to be solved.

We first derive uncertainty relations, beginning the usual procedure familiar
from textbooks on quantum mechanics. We pick a pair $(\hat{x}_i,\hat{x}_j)$ of
basic operators, assumed self-adjoint as noted before. Starting with a generic
state $\psi$ in a Hilbert-space representation of the basic algebra, we
introduce three new states $v_{x_i}:=\widehat{\Delta x_i}\psi$,
$w_{x_j}:=\widehat{\Delta x_j}\psi$ and $u_{x_i,x_j}:= w_{x_j}-(v_{x_i}\cdot
w_{x_j}/||v_{x_i}||^2) v_{x_i}$. The uncertainty relation for our pair of
operators then follows from
\begin{equation}
 0\leq ||u_{x_i,x_j}||^2= ||w_{x_j}||^2- \frac{|v_{x_i}\cdot
   w_{x_j}|^2}{||v_{x_i}||^2}\,, 
\end{equation}
with saturation if and only if $u_{x_i,x_j}=0$.

Inserting our specific expressions for $u_{x_i}$ and $w_{x_j}$ in terms of
$\psi$, we can express each term in the Schwarz inequality
$||v_{x_i}||^2||w_{x_j}||^2\geq |v_{x_i}\cdot w_{x_j}|^2$ in terms of
moments. We easily obtain $||v_{x_i}||^2=\Delta(x_i^2)$ and
$||w_{x_j}||^2=\Delta(x_j^2)$ and, with a little more re-ordering work,
\[
 |v_{x_i}\cdot w_{x_j}|^2= |\langle (\hat{x}_i-x_i)(\hat{x}_j-x_j)\rangle|^2=
 |\Delta(x_ix_j)+ {\textstyle\frac{1}{2}} i\hbar \epsilon_{ij}{}^kx_k|^2=
 \Delta(x_ix_j)^2+ \frac{\hbar^2}{4}(\epsilon_{ij}{}^kx_k)^2\,.
\]
Uncertainty relations
\begin{equation} \label{Uncertx}
 \Delta (x_i^2)\Delta (x_j^2)-\Delta(x_ix_j)^2\geq \frac{\hbar^2}{4}
 (\epsilon_{ij}{}^kx_k)^2
\end{equation}
in standard form then follow.

\subsection{Higher orders}

Higher-order moments are restricted by higher-order uncertainty relations. We
can derive them by using non-linear polynomials in the $\widehat{\Delta x_i}$
to define states $v_{{\rm pol}_1}:=\widehat{{\rm pol}_1}\psi$ and $w_{{\rm
    pol}_2}:=\widehat{{\rm pol}_2}\psi$ and proceeding as before. Without loss
of generality, we require $\widehat{{\rm pol}_{1/2}}$ to be
Weyl-ordered. Unlike (\ref{Uncertx}), these higher-order relations mix moments
of different orders. (For recent work on higher-order uncertainty relations
with canonical basic operators, see \cite{ClassMoments}.)

The first ones beyond (\ref{Uncertx}), for instance, involve moments
of second, third and fourth order, obtained when ${\rm pol}_1$ is linear and
${\rm pol}_2$ quadratic (or vice versa). Using the relations in the appendix,
we compute 
\begin{eqnarray}
 ||w_{x_jx_k}||^2 &=& \Delta(x_j^2x_k^2)\\
&&-\frac{1}{6} \hbar^2 \left(
 \epsilon_{jk}{}^l \epsilon_{jl}{}^m \Delta(x_kx_m) -
\epsilon_{jk}{}^l \epsilon_{kl}{}^m\Delta(x_jx_m)\right)- \frac{3}{4}\hbar^2 
\epsilon_{jk}{}^l
\epsilon_{jk}{}^m \Delta(x_lx_m) \nonumber
\end{eqnarray}
(no sum over repeated lower indices) and 
\begin{eqnarray}
 |\langle v_{x_i},w_{x_jx_k}\rangle|^2 &=& \left(\Delta(x_ix_jx_k)-
   \frac{1}{12}\hbar^2\left(\epsilon_{ik}{}^l \epsilon_{lj}{}^m+
     \epsilon_{ij}{}^l \epsilon_{lk}{}^m\right)x_m\right)^2\\
 && + \frac{1}{4}\hbar^2\left(   \epsilon_{ij}{}^l\Delta(x_kx_l)+
   \epsilon_{ik}{}^l\Delta(x_jx_l)\right)^2\,. \nonumber
\end{eqnarray}

If we only consider terms of lowest order (six), this third-order uncertainty
relation becomes
\begin{equation} \label{UncertThird}
 \Delta(x_j^2x_k^2)\Delta(x_i^2)- \Delta(x_ix_jx_k)^2\geq 0.
\end{equation}
Unless there are third-order correlations between the basic variables, the
only implication is that fourth-order moments of the form $\Delta(x_j^2x_k^2)$
must be positive, which already follows from their definition. 

The next order of uncertainty relations is more interesting and bounds
fourth-order moments by a positive number. Defining $\vec{\hat{x}}{}^{\vec{i}}
= \left(\hat{x}_1^{i_1} \hat{x}_2^{i_2} \ldots \hat{x}_M^{i_M}\right)_{\rm
  Weyl-ordered}$, in general, the leading-order contribution to the
uncertainty relation implied by $\widehat{{\rm
    pol}_1}=\vec{\hat{x}}{}^{\vec{i}}$ and $\widehat{{\rm
    pol}_2}=\vec{\hat{x}}{}^{\vec{j}}$ is of the form
\begin{equation} \label{UncertGen}
 \Delta(\vec{x}^{2\vec{i}})\Delta(\vec{x}^{2\vec{j}})-
 \Delta(\vec{x}^{2\vec{i}+2\vec{j}})\geq U
\end{equation}
where $U$ follows from the squared imaginary part of
$\langle\vec{\hat{x}}{}^{\vec{i}}\vec{\hat{x}}{}^{\vec{j}}\rangle-
\Delta(\vec{x}^{\vec{i}+\vec{j}})$. The leading order is obtained if exactly
one commutator is applied in the re-ordering required to bring
$\langle\vec{\hat{x}}{}^{\vec{i}}\vec{\hat{x}}{}^{\vec{j}}\rangle$ into
$\Delta(\vec{x}^{\vec{i}+\vec{j}})$. It has two contributions, one of degree
$2(|\vec{i}|+|\vec{j}|+1)$ (after taking the square) from the $\widehat{\Delta
  x_k}$-term in (\ref{DeltaComm}), and one of order $2(|\vec{i}|+|\vec{j}|)$
from the $x_k$-term. The latter is always of the same order as the leading
contribution on the left-hand side of (\ref{UncertGen}), except when it
happens to vanish. It always vanishes for third-order uncertainty relations
(\ref{UncertThird}) because it contains only the vanishing
$\Delta(\vec{x}^{\vec{i}})$ with $|\vec{i}|=1$. For higher orders, however,
$U$ in (\ref{UncertGen}) is non-zero to leading order, so that the familiar
form of uncertainty relations is obtained, with the right-hand side non-zero
and proportional to $\hbar^2$.

\subsection{Relations for group coherent states}

Group coherent states obey Casimir conditions and saturate uncertainty
relations. We will now explore the interplay of these different conditions,
focusing on the case of a 3-dimensional semisimple Lie group of rank one. In
particular, the counting problem, treated in the previous section for a
Casimir condition irrespective of uncertainty relation, is again of
relevance. On the unrestriced state space, ignoring the Casimir condition, we
obtain an independent uncertainty relation (\ref{Uncertx}) for every pair of
generators $\hat{x}_i$. When the Casimir condition is imposed, some degrees of
freedom are removed, and we should expect a smaller number of independent
uncertainty relations. For instance, for a 3-dimensional Lie group of rank
one, we start with three second-order uncertainty relations, one for every
pair among three basic operators, but should have only one independent one for
the 2-dimensional phase space left after the Casimir condition is
imposed. This counting problem is difficult to analyze at arbitrary orders,
but it is instructive to have a look at the leading orders of second-order
moments that feature in (\ref{Uncertx}).

Under the present assumptions, we have a quadratic Casimir
$C=k^{ij}x_ix_j-c=0$ with a constant $c$, using the Killing metric
$k^{ij}$. To lowest order, such a condition implies that
\begin{equation}\label{CbarKill}
\bar{C}_{\vec{i}}= 2x_ik^{ij}\Delta(\vec{x}^{\vec{i}} x_j)=0\,,
\end{equation}
using (\ref{Cbar}).  For higher-order contributions, we must perform
more-explicit calculations keeping track of all necessary re-orderings. It is
then useful to write the corresponding Casimir operator in terms of
$\widehat{\Delta x_i}$:
\begin{eqnarray*} 
  \hat{C}&=&k^{ij}\hat{x}_i\hat{x}_j-c\hat{\mathbf{1}}= 
 k^{ij}(\widehat{\Delta x_i}+x_i)
  (\widehat{\Delta x_j}+x_j)-c\hat{\mathbf{1}}\\
  &=& k^{ij} \widehat{\Delta x_i}\widehat{\Delta x_j}+ 2k^{ij} x_i
  \widehat{\Delta x_j}+ k^{ij} x_ix_j-c\hat{\mathbf{1}}\,.
\end{eqnarray*}
The effective Casimir condition then reads
\begin{equation} \label{CasEff}
 \langle\hat{C}\rangle= k^{ij} \Delta(x_ix_j)+ k^{ij}x_ix_j-c=0\,,
\end{equation}
and can be used to simplify $\hat{C}$ ``on-shell,'' that is when the basic
Casimir condition holds:
\begin{equation} \label{CApp}
 \hat{C}\approx k^{ij} (\widehat{\Delta x_i}\widehat{\Delta
   x_j}-\Delta(x_ix_j))+ 2k^{ij} x_i \widehat{\Delta x_j}\,.
\end{equation}
In this form, the condition is useful for higher-order effective constraints
because the lower-order ones will already be taken care of. Moreover,
(\ref{CasEff}) does not explicitly depend on $c$, so that
representation-independent relations between expectation values and moments of
different orders will be obtained. (The actual values of moments depend on $c$
and the representation once specific solutions to (\ref{CasEff}) are used.)

The next-order conditions are obtained from $\frac{1}{2}\langle
\widehat{\Delta x_k}\hat{C}+ \hat{C}\widehat{\Delta x_k}\rangle$, in which we
directly write the symmetric ordering (and use $\widehat{\Delta x_k}$ because
the $x_k$-terms subtracted from $\hat{x}_k$ would just multiply lower-order
constraints). However, not all terms in
$k^{ij}(\hat{x}_k\hat{x}_i\hat{x}_j+\hat{x}_i\hat{x}_j\hat{x}_k)$ are ordered
totally symmetric and give rise to moments as defined in
(\ref{Momentsx}). To change the ordering in this cubic case, we make use of the
identity
\begin{equation}
 \frac{1}{2}(\hat{A}\hat{B}\hat{D}+\hat{B}\hat{D}\hat{A})=
 (\hat{A}\hat{B}\hat{D})_{\rm Weyl-ordered}+
 \frac{1}{12}([[\hat{A},\hat{B}],\hat{D}]+ 
 [[\hat{A},\hat{D}],\hat{B}])+ \frac{1}{4}\{\hat{A},[\hat{B},\hat{D}]\}
\end{equation}
and write
\begin{eqnarray*}
 \frac{1}{2}\left(
\widehat{\Delta x_k}\hat{C}+ \hat{C}\widehat{\Delta x_k}\right)&=&
k^{ij}(\widehat{\Delta x_i}\widehat{\Delta x_j}\widehat{\Delta x_k})_{\rm
  Weyl-ordered}\\
&&+ k^{ij} \left(\frac{1}{6}[[\widehat{\Delta x_k},\widehat{\Delta x_i}],
\widehat{\Delta x_j}]- \Delta(x_ix_j)\widehat{\Delta x_k}\right)+2
k^{ij}x_i(\widehat{\Delta x_j}\widehat{\Delta x_k})_{\rm Weyl-ordered}\,.
\end{eqnarray*}
Its expectation value gives the third-order Casimir condition
\begin{equation}
 \frac{1}{2}\langle
\widehat{\Delta x_k}\hat{C}+ \hat{C}\widehat{\Delta x_k}\rangle=
k^{ij}\Delta(x_ix_jx_k)- \frac{\hbar^2}{6}
k^{ij}\epsilon_{ki}{}^l\epsilon_{lj}{}^m x_m + 2k^{ij}x_i \Delta(x_jx_k)\approx0\,.
\end{equation}
Even though the condition is derived from third-order operators, its leading
term (the last one in the equation) is of second order because the Casimir
operator does not depend just on $\widehat{\Delta x_i}$. If moments (and
explicit factors of $\hbar$) of higher than second order are ignored, the
third-order Casimir condition restricts second-order moments by
\begin{equation} \label{ThirdCasimir}
 x^j\Delta(x_jx_k)\approx 0\,,
\end{equation}
where we raise the index using the Killing metric. This result is clearly a
special case of (\ref{CbarKill}).

To the next order, we obtain from $\langle(\widehat{\Delta x_k}\widehat{\Delta
  x_l})_{\rm Weyl-ordered}\hat{C}\rangle=0$ the relation
\begin{eqnarray} \label{FourthCasimir}
 &&2x^i\Delta(x_ix_kx_l)+ k^{ij} \left(\Delta(x_ix_jx_kx_l)-
 \Delta(x_ix_x)\Delta(x_kx_l)\right)\\
&&+\frac{1}{6}\hbar^2 k^{ij} \left(3\epsilon_{i(k}{}^m \epsilon_{l)j}{}^n
  \Delta(x_mx_n)+ 2\epsilon_{i(k}{}^m \epsilon_{l)m}{}^n \Delta(x_jx_n)+2
  \epsilon_{mj}{}^n \epsilon_{i(k}{}^m \Delta(x_{l)}x_n)\right)\nonumber\\
&& -\frac{1}{6}\hbar^2 k^{ij} \epsilon_{ki}{}^m \epsilon_{lj}{}^n x_mx_n
 \approx0\,. \nonumber
\end{eqnarray}
For totally antisymmetric $\epsilon_{ijk}$, the second line simplifies to
$-\frac{5}{6} \hbar^2 \Delta(x_lx_k)-\frac{1}{6} \hbar^2\delta_{lk}
\Delta(x^ix_i)$, and the last line to
$-\frac{1}{6}\hbar^2(\delta_{lk}x^ix_i-x_lx_k)$.

\subsection{Interplay of Casimir conditions and uncertainty relations}

With the more-explicit form of effective Casimir conditions found in the
preceding subsection, we can see how the number of uncertainty relations is
reduced. As before, we assume a 3-dimensional Lie algebra with three
uncertainty relations (\ref{Uncertx}). We start with the relation obtained for
the pair $(\hat{x}_1,\hat{x}_2)$ and use (\ref{ThirdCasimir}) to bring it to
the form of one of the others. To this end, we multiply the left-hand side of
(\ref{Uncertx}) with $(x^1)^2$ and rewrite using
\begin{eqnarray*}
x^1\Delta(x_1^2) &=& -x^2\Delta(x_2x_1)-x^3\Delta(x_3x_1)\\
x^1\Delta(x_1x_2) &=& -x^2\Delta(x_2^2)-x^3\Delta(x_3x_2)\\
x^1\Delta(x_1x_3) &=& -x^2\Delta(x_2x_3)-x^3\Delta(x_3^2)
\end{eqnarray*}
several times.
We obtain
\begin{eqnarray*}
 (x^1)^2 (\Delta(x_1^2)\Delta(x_2^2)-\Delta(x_1x_2)^2) &=&
 ((x^2)^2\Delta(x_2^2)+2x^2x^3\Delta(x_2x_3)+(x^3)^2\Delta(x_3^2))
 \Delta(x_2^2)\\
&& -(x^2\Delta(x_2^2)+x^3\Delta(x_2x_3))^2\\
&=& (x^3)^2(\Delta(x_2^2)\Delta(x_3^2)-\Delta(x_2x_3)^2)\,,
\end{eqnarray*}
just as needed for the left-hand side of the uncertainty relation belonging to
the pair $(\hat{x}_2,\hat{x}_3)$. 

On the right-hand side, we have $x^1\epsilon_{12}{}^kx_k=
x^1\epsilon_{123}x^3=- x^3\epsilon_{32k}x^k$, using the behavior of structure
constants for rank-1 semisimple Lie algebras. Also the right-hand sides will
then match, showing that the uncertainty relations for $(\hat{x}_1,\hat{x}_2)$
and $(\hat{x}_2,\hat{x}_3)$ are not independent. Similarly, one can show that
the one for $(\hat{x}_1,\hat{x}_3)$ is not independent either, demonstrating
that only one independent uncertainty relation is left after the Casimir
condition is imposed.

Including higher orders in the Casimir constraints, we have 
\begin{eqnarray*}
x^1\Delta(x_1^2) &=& -x^2\Delta(x_2x_1)-x^3\Delta(x_3x_1)-
\frac{1}{2}\Delta(x^ix_ix_1)+ \frac{1}{6}\hbar^2 x_1\\
x^1\Delta(x_1x_2) &=& -x^2\Delta(x_2^2)-x^3\Delta(x_3x_2)-
\frac{1}{2}\Delta(x^ix_ix_2)+ \frac{1}{6}\hbar^2 x_2\\
x^1\Delta(x_1x_3) &=& -x^2\Delta(x_2x_3)-x^3\Delta(x_3^2)-
\frac{1}{2}\Delta(x^ix_ix_3)+ \frac{1}{6}\hbar^2 x_3
\end{eqnarray*}
and
\begin{eqnarray*}
 (x^1)^2 (\Delta(x_1^2)\Delta(x_2^2)-\Delta(x_1x_2)^2)
&=& (x^3)^2(\Delta(x_2^2)\Delta(x_3^2)-\Delta(x_2x_3)^2)\\
 &&-\frac{1}{2}x^j\Delta(x^ix_ix_j) \Delta(x_2^2)\\
&&+ x^3\left(
   \Delta(x^ix_ix_3)\Delta(x_2^2)- \Delta(x^ix_ix_2)\Delta(x_2x_3)\right)\\
&& +\frac{1}{6}\hbar^2 \left((x^1x_1+x^2x_2-x^3x_3) \Delta(x_2^2)+
  2x^3x_2\Delta(x_2x_3)\right) \\
&&-\frac{1}{4}\Delta(x^ix_ix_2)^2+ \frac{1}{6}\hbar^2 x_2\Delta(x^ix_ix_2)-
\frac{1}{36}\hbar^4 x_2^2 \,.
\end{eqnarray*}
In the second line one can use (\ref{FourthCasimir}), which brings in
fourth-order moments.  This shows that the interplay of different uncertainty
relations subject to Casimir conditions is more complicated and mixes the
orders. A more systematic description of higher-order uncertainty relations
would be useful.

\section{Example: Reality conditions of harmonic loop quantum cosmology}

As a detailed application, we consider, as introduced in Sec.~\ref{s:EffCas},
basic operators satisfying the ${\rm sl}(2,{\mathbb R})$ algebra
\begin{equation}\label{comm}
  [\hat{V},\hat{J}]=\hbar\hat{J}\quad,\quad {}
  [\hat{V},\hat{J}^{\dagger}]=-\hbar\hat{J}^{\dagger}\quad,\quad {}
  [\hat{J},\hat{J}^{\dagger}]=-2\hbar(\hat{V}+\hbar/2)
\end{equation}
subject to the condition $\hat{J}\hat{J}^{\dagger}=\hat{V}^2$. (This example
is not exactly of the form discussed previously, owing to the central
extension by $\hbar/2$. However, as this central extension is trivial, the
same results hold true. One may absorb $\hbar/2$ in the definition of
$\hat{V}$, but we will keep this term explicitly since it contributes to some
quantum corrections.) For expectation values, the operator identity
$\hat{C}=\hat{J}\hat{J}^{\dagger}-\hat{V}^2$ implies the equation
\begin{equation} \label{reality}
 |J|^2-(V+\hbar/2)^2= (\Delta V)^2-\Delta(JJ^*)+\frac{1}{4}\hbar^2
\end{equation}
relating expectation values to fluctuations of $\hat{V}$, $\hat{J}$ and
$\hat{J}^{\dagger}$. Alternatively, we may write the condition as
\begin{equation}
 J_+^2+J_-^2-V^2= (\Delta V)^2-(\Delta J_+)^2-(\Delta
 J_-)^2+\frac{1}{2} \hbar(2V+\hbar)
\end{equation} 
for moments of self-adjoint operators $\hat{V}$, $\hat{J}_+=
\frac{1}{2}(\hat{J}+\hat{J}^{\dagger})$ and $\hat{J}_-=
-\frac{1}{2}i(\hat{J}-\hat{J}^{\dagger})$.

Equation (\ref{reality}) reduces the number of degrees of freedom
contained in the expectation values back to the usual canonical value
of two.  In this counting, the fluctuations on the right of
(\ref{reality}) are considered fixed.  Independent conditions for
fluctuations and higher moments arise at higher order by considering
operators
\begin{equation} \label{realityOp}
\left((\Delta \hat{V})^i (\Delta \hat{J}_+)^j (\Delta\hat{J}_-)^k\right)_{\rm
  Weyl}  (\hat{J}_+^2+\hat{J}_-^2-\hat{V}^2)\quad
\mbox{with}\quad i+j+k>0\,,
\end{equation}
whose expectation values vanish.  In terms of $\Delta$-operators, the reality
condition reads
\begin{eqnarray*}
 \hat{C}&=& \widehat{\Delta J_+}^2+\widehat{\Delta J_-}^2- \widehat{\Delta V}^2+
 2J_+\widehat{\Delta J_+}+2 J_-\widehat{\Delta J_-}- (2V+\hbar)\widehat{\Delta
   V}\\
&&+ J_+^2+J_-^2-(V+\hbar/2)^2-\hbar^2/4\\
&\approx& \widehat{\Delta J_+}^2-\Delta(J_+^2) +\widehat{\Delta
  J_-}^2-\Delta(J_-^2) - \widehat{\Delta V}^2+\Delta(V^2)\\
&&+ 2J_+\widehat{\Delta J_+}+2 J_-\widehat{\Delta J_-}-
(2V+\hbar)\widehat{\Delta V} 
\end{eqnarray*}
analogous to (\ref{CApp}). For symmetric reorderings of $\Delta$-operators, we
then use the relations
\begin{eqnarray}
 [\widehat{\Delta V},\widehat{\Delta J_+}]&=& i\hbar(\widehat{\Delta
   J_-}+J_-)\\{} 
 [\widehat{\Delta V},\widehat{\Delta J_-}]&=& -i\hbar(\widehat{\Delta
   J_+}+J_+)\\{} 
[\widehat{\Delta J_+},\widehat{\Delta J_-}]&=&
 -i\hbar(\widehat{\Delta V}+V+\hbar/2)\,.
\end{eqnarray}

For instance, the third-order moments
\begin{eqnarray}
\Delta(VJ_{\pm}^2) &\equiv&
\left\langle ( \hat{V} - V ) ( \hat{J}_{\pm}- J_{\pm})^2
 \right\rangle_{\rm Weyl-ordered}  \\
\Delta(V^3) &\equiv&\left\langle ( \hat{V}  - V)^3  \right\rangle 
\end{eqnarray} 
appear in the reality condition
\begin{eqnarray*}
0&=&\Delta(VJ_+^2)+\Delta(VJ_-^2)-\Delta(V^3)\\
&&+2J_+\Delta(VJ_+)+ 2J_-\Delta(VJ_-)- (2V+\hbar)\Delta(V^2)\\
&&-\frac{1}{6}\hbar^2(2V+\hbar)
\end{eqnarray*}
following from the vanishing expectation value of
$\Delta \hat{V}(\hat{J}_+^2+\hat{J}_-^2-\hat{V}^2)$. Similarly,
\begin{eqnarray*}
 0&=& \Delta(J_+^3)+\Delta(J_+J_-^2)-\Delta(J_+V^2)\\
&&+2J_+\Delta(J_+^2)+2J_-\Delta(J_+J_-)-2V\Delta(VJ_+)-\hbar \Delta(VJ_+)
 -\frac{\hbar^2}{6}J_+\,,\\
 0&=& \Delta(J_-^3)+\Delta(J_+^2J_-)-\Delta(J_-V^2)\\
&&+2J_-\Delta(J_-^2)+2J_+\Delta(J_+J_-)-2V\Delta(VJ_-)-\hbar \Delta(VJ_-)
-\frac{\hbar^2}{6}J_-\,.
\end{eqnarray*}

Third-order reality conditions restrict semiclassical second-order moments.
To leading order in $\hbar$ (keeping only the ``central charge'' $\hbar/2$ as
a higher-order contribution to $V$),
\begin{eqnarray}
 (V+\hbar/2)\Delta(V^2) &=& J_+\Delta(VJ_+)+J_-\Delta(VJ_-)\\
 (V+\hbar/2)\Delta(VJ_+) &=& J_+\Delta(J_+^2)+J_-\Delta(J_+J_-)\\
 (V+\hbar/2)\Delta(VJ_-) &=& J_-\Delta(J_-^2)+J_+\Delta(J_+J_-)\,.
\end{eqnarray}
In terms of moments of complex variables, as derived in \cite{HighDens} and
used crucially to restrict initial values of moments for numerical solutions
of equations of motion, this reads
\begin{equation}  \label{realitySecond1}
 (V+\hbar/2)(\Delta V)^2={\rm
   Re}(J^*\Delta(VJ))=
{\rm Re} J {\rm Re} \Delta(VJ) + {\rm Im} J {\rm Im} \Delta(VJ)\,.
\end{equation}
The remaining third-order relations imply
\begin{eqnarray} \label{realitySecond2}
 (V+\hbar/2) {\rm Re} \Delta(VJ) &=& \frac{1}{2}\left( {\rm
     Re} J {\rm Re} (\Delta J)^2 + 
 {\rm Im} J {\rm Im} (\Delta J)^2 + {\rm Re}
 J \Delta(JJ^*)
\right)\,, \nonumber \\
 (V+\hbar/2){\rm Im} \Delta(VJ) &=& \frac{1}{2}\left( {\rm
     Re} J  {\rm Im} (\Delta J)^2 - 
{\rm Im} J {\rm Re} (\Delta J)^2 + {\rm Im}
J  \Delta(JJ^*)
\right)\,.
\end{eqnarray}
By our general considerations, none of the higher-order conditions restrict
second-order moments further. (The lowest-order term given by (\ref{CbarKill})
is of order three or higher for Casimir conditions of order four or higher.)
Initially, we have six second-order moments.  By higher-order reality
conditions, they are subject to three further conditions, leaving three
degrees of freedom as expected for two fluctuations and one correlation.

\subsection{Uncertainty relations and existence of coherent states}

For the pairs $(\hat{V},\hat{J}_+)$, $(\hat{V},\hat{J}_-)$ and
$(\hat{J}_+,\hat{J}_-)$ of self-adjoint operators, we obtain
from (\ref{Uncertx}):
\begin{eqnarray}
 (\Delta V)^2 (\Delta J_+)^2- \Delta(VJ_+)^2
 &\geq& \frac{1}{4} \hbar^2J_-^2\label{uncert1}\\
 (\Delta V)^2(\Delta J_-)^2- \Delta(VJ_-)^2
  &\geq& \frac{1}{4}\hbar^2J_+^2\label{uncert2}\\
 (\Delta J_+)^2(\Delta J_-)^2-
 \Delta(J_+J_-)^2
 &\geq& \frac{1}{4}\hbar^2(V+\hbar/2)^2\,. \label{uncert3}
\end{eqnarray}
Our previous counting argument has shown that one canonical pair and its
moments are left after the Casimir condition is imposed.  As in the general
case of Casimir conditions, the reality conditions ensure that two of the
uncertainty relations are indeed equivalent to the third one.

A state saturating the uncertainty relation and obeying the Casimir condition
is a coherent state for the group that provides the phase space. If we solve
the saturated uncertainty relation (or its higher-order analogs) that remains
after the Casimir condition has been used, we obtain moments that could belong
to a coherent state, but it is not obvious that there is an actual
(normalizable) wave function for them. In fact, constructing explicit wave
functions for coherent states can be a complicated procedure. Fortunately, in
a concrete example, one can show, without constructing the actual wave
function, that there is always a wave function that produces ``coherent''
moments obtained by solving saturated uncertainty relations.

To show this, we begin in the standard way used in quantum mechanics, where a
Gaussian is obtained as the unique wave function saturating the canonical
uncertainty relation. For our two basic operators $\hat{V}$ and $\hat{J}_+$
and some state $\psi$ we introduce, as before, two states $v:=(\hat{V}-V)\psi$
and $w:=(\hat{J}_+-J_+)\psi$, with expectation values taken in the same state
$\psi$. From standard arguments it then follows that the $(V,J_+)$-uncertainty
relation is saturated if and only if $u:=w-(v\cdot w/||v||^2)v$ vanishes. In our
example, this equation reads
\begin{equation} \label{psiSat}
 \hat{J}_+\psi-
 \frac{\Delta(VJ_+)+\frac{1}{2}\langle[\hat{V},\hat{J}_+]\rangle}{(\Delta
 V)^2} \hat{V}\psi+
 \left(\frac{\Delta(VJ_+)+\frac{1}{2}\langle[\hat{V},\hat{J}_+]\rangle}{(\Delta
 V)^2} V-J_+\right) \psi=0\,.
\end{equation}

We may represent states as wave functions $\psi=\sum_n \psi_n|n\rangle$ in
terms of $\hat{V}$-eigenstates $|n\rangle$ with $\hat{V}|n\rangle=n|n\rangle$
(assuming the discrete series of representations, which is relevant for the
quantum-cosmological application). On these eigenstates, using the realization
$\hat{J}=\exp(-i\hat{P})$ in terms of canonical operators $(\hat{V},\hat{P})$,
$\hat{J}|n\rangle=(n+1)|n+1\rangle$ acts like the product of $\hat{V}$ with a
raising operator, and the Hermitian combination as
\[
 \hat{J}_+|n\rangle= \frac{1}{2}\left((n+1)|n+1\rangle+ n|n-1\rangle\right)\,.
\]
The saturation equation (\ref{psiSat}) for a wave function $\psi$ is therefore
equivalent to a difference equation
\begin{equation} \label{Diff}
 \frac{1}{2}\left(n\psi_{n-1}+(n+1)\psi_{n+1}\right) 
-\alpha n\psi_n+\beta\psi_n=0
\end{equation}
where we denoted the coefficients in (\ref{psiSat}) as 
\begin{equation} \label{alpha}
 \alpha=\frac{\Delta(VJ_+)+
 \frac{1}{2}\langle[\hat{V},\hat{J}_+]\rangle}{(\Delta
 V)^2} =\frac{\Delta(VJ_+)+\frac{1}{2}i\hbar J_-}{(\Delta V)^2}
\end{equation}
and 
\[
 \beta=\alpha V-J_+\,.
\]
Since expectation values and moments in $\alpha$ and $\beta$ themselves depend
on $\psi$, and therefore on all $\psi_n$, the system of coupled equations
defined by (\ref{Diff}) is non-linear and difficult to deal with exactly.  For
our purposes, however, it is sufficient to analyze the asymptotic behavior for
$n\to\pm\infty$, which is feasible.

The range of $n$ includes all integers, and therefore suitable
fall-off conditions must be satisfied at both ends for a normalizable
state in $\ell^2$. However, it is sufficient to show that (\ref{Diff})
has at least one normalizable solution for large positive and
negative values of $n$, respectively, because one can always patch
together a normalizable solution on $(-\infty,-1]$ with one on
$(1,\infty]$ by choosing $\psi_0$ so that (\ref{Diff}) holds for
$n=0$. 

For the existence of normalizable solutions at large $n$, we consider the
difference equation $\psi_{n+1}-2\alpha\psi_n+ \psi_{n-1}=0$ with constant
coefficients. Its solutions are $\psi^{\pm}_n=k_{\pm}^n$ with
$k_{\pm}=\alpha\pm\sqrt{\alpha^2-1}$. Solutions $\psi_{\pm}$ are normalizable
if $|k_{\pm}|<1$.  These two numbers satisfy the relation $k_+k_-=1$, and
therefore there can be at most one normalizable solution. There is no
normalizable solution if and only if both $k_{\pm}$ lie on the unit circle, in
which case $k_+=k_-^*$. The latter condition can be fulfilled only if $\alpha$
is real with $|\alpha|\leq 1$. In all other cases, a normalizable solution
exists and we are guaranteed to have a state saturating the uncertainty
relation. Here, (\ref{alpha}) is not real unless $J_-=0$, which is generic
enough to conclude the existence of coherent states. Solving effective
equations for moments produces their quantum parameters, even if an explicit
wave function is unknown. In the present example, we can use these results to
extend \cite{GroupLQC}, where wave functions for ${\rm sl}(2,{\mathbb
  R})$-coherent states were obtained for small correlations. The moments of
group coherent states found here then allow us to address cosmic
forgetfulness, for which potentially large correlations and squeezing are
important \cite{BounceCohStates}.

\subsection{Moments of dynamical coherent states}

For simple Hamiltonian operators, one can find dynamical coherent states that
saturate the uncertainty relation at all times.  Harmonic loop quantum
cosmology \cite{BouncePert}, based on the non-canonical algebra (\ref{comm})
with the Hamiltonian $\hat{H}=\hat{J}_-$, provides such an example. Exact
solutions for expectation values and moments in arbitrary states can first be
found, and then restricted to those that saturate the uncertainty
relation. The dynamical equations are slightly more compact if we use complex
variables $(V,J,J^*)$ and the corresponding moments.

Evolution in some time parameter $\lambda$ is generated by $\hat{H}$. As per
Ehrenfest's equations, expectation values obey
\begin{eqnarray} \label{Vdot}
 \frac{{\rm d}}{{\rm d} \lambda} V&=&
 \frac{1}{i\hbar}\langle[\hat{V},\hat{H}]\rangle=
 -\frac{1}{2}(J+J^*)\\
 \frac{{\rm d}}{{\rm d} \lambda} J&=&
 \frac{1}{i\hbar}\langle[\hat{J},\hat{H}]\rangle=
 -(V+\hbar/2)=\frac{{\rm d}}{{\rm d}
 \lambda}J^*\,, \label{Jdot}
\end{eqnarray}
solved by 
\begin{equation}\label{VJ}
 V(\lambda)+\hbar/2=A \cosh(\lambda-\lambda_0)\quad,\quad
 J(\lambda)=-A\sinh(\lambda-\lambda_0)+iH
\end{equation}
since ${\rm Im}J=J_-=\langle\hat{H}\rangle=:H$, and with integration constants
$A$ and $\lambda_0$. Second-order moments satisfy the equations of motion
\cite{BounceCohStates}
\begin{eqnarray}
 \frac{{\rm d}}{{\rm d}\lambda}(\Delta V)^2 &=& -\Delta(VJ)-\Delta(VJ^*) 
  \label{GVVdot}\\ 
\frac{{\rm d}}{{\rm d}\lambda}(\Delta J)^2 &=& -2\Delta(VJ)\quad,\quad
 \frac{{\rm d}}{{\rm d}\lambda}(\Delta J^*)^2 = -2\Delta(VJ^*)\\ 
 \frac{{\rm d}}{{\rm d}\lambda}\Delta(VJ) &=&
 -\frac{1}{2} (\Delta J)^2-\frac{1}{2}\Delta(JJ^*)-(\Delta V)^2\\
 \frac{{\rm d}}{{\rm d}\lambda}\Delta(VJ^*) &=& -\frac{1}{2} (\Delta J^*)^2-
 \frac{1}{2}\Delta(JJ^*)-(\Delta V)^2\\ 
\frac{{\rm d}}{{\rm d}\lambda}\Delta(JJ^*) &=&
 -\Delta(VJ)-\Delta(VJ^*) \label{GJbarJdot}
\end{eqnarray}
solved by
\begin{eqnarray*}
 (\Delta V)^2(\lambda) &=& \frac{1}{2}(c_3e^{-2\lambda}+c_4e^{2\lambda})- 
\frac{1}{4}(c_1+c_2)\\
 (\Delta J)^2(\lambda) &=& \frac{1}{2}(c_3e^{-2\lambda}+c_4e^{2\lambda})+
\frac{1}{4}(3c_2-c_1)-
 i(c_5e^{\lambda}-c_6e^{-\lambda})\\
 (\Delta J^*)^2(\lambda) &=&
 \frac{1}{2}(c_3e^{-2\lambda}+c_4e^{2\lambda})+ 
\frac{1}{4}(3c_2-c_1)+
 i(c_5e^{\lambda}-c_6e^{-\lambda})\\
 \Delta(VJ)(\lambda) &=& \frac{1}{2}(c_3e^{-2\lambda}-c_4e^{2\lambda})+ 
\frac{i}{2}(c_5e^{\lambda}+c_6e^{-\lambda})\\
 \Delta(VJ^*)(\lambda) &=& \frac{1}{2}(c_3e^{-2\lambda}-c_4e^{2\lambda})- 
\frac{i}{2}(c_5e^{\lambda}+c_6e^{-\lambda})\\
 \Delta(JJ^*)(\lambda) &=& \frac{1}{2}(c_3e^{-2\lambda}+c_4e^{2\lambda})+
 \frac{1}{4}(3c_1-c_2)\,.
\end{eqnarray*}
These equations and some of the following derivations can be found in
\cite{BounceCohStates,Harmonic}. We list them here because we will be able to
generalize them in the next subsection, for which we have to refer to the
older results.

Here, $c_1=-(\Delta V)^2+\Delta(JJ^*)$ which by the reality condition
(\ref{reality}) equals 
\begin{equation} \label{c1}
 c_1=A^2-H^2
\end{equation}
in terms of $H=\langle\hat{H}\rangle$ and the integration constant $A$ in the
solutions $V(\lambda)$ and $J(\lambda)$ in (\ref{VJ}). Using
$\hat{H}=-\frac{1}{2}i(\hat{J}-\hat{J}^{\dagger})$ we obtain
\begin{equation}
 (\Delta H)^2 = -\frac{1}{4}\left((\Delta J)^2-2\Delta(JJ^*)+
   (\Delta J^*)^2\right)
= \frac{1}{2}(c_1-c_2) \label{c1c2}\,.
\end{equation}
The remaining constants are subject to further conditions from reality and
uncertainty bounds. From (\ref{uncert1}), requiring saturation,
\begin{equation} \label{Uncert}
 4c_3c_4 = H^2\hbar^2+\frac{1}{4}(c_1+c_2)^2 \,.
\end{equation}
Additional conditions follow from the other uncertainty relations, but
they are equivalent once reality conditions are imposed.

The second-order reality conditions (\ref{realitySecond1}) and
(\ref{realitySecond2}), evaluated for the explicit solutions, are
$\lambda$-dependent.  Comparing coefficients of different powers of
$e^{\lambda}$, we obtain three independent equations
\begin{eqnarray}
 c_5 =
 \frac{A}{2H}\left(2c_4-{\textstyle\frac{1}{2}}(c_1+c_2)\right)\quad &,&\quad
 c_6 =
 \frac{A}{2H}\left(2c_3-{\textstyle\frac{1}{2}}(c_1+c_2)\right) \label{c5c6}\\
c_5+c_6 &=& \frac{H}{A}(c_1-c_2)\,. \label{c5plusc6}
\end{eqnarray}
We use (\ref{c5plusc6}) to eliminate $c_5$ and $c_6$ from the
sum of the two equations in (\ref{c5c6}),
\begin{equation} \label{c3plusc4}
 c_3+c_4= \frac{H}{A}(c_5+c_6)+\frac{1}{2}(c_1+c_2)= \frac{H^2}{A^2}
 (c_1-c_2)+\frac{1}{2}(c_1+c_2)\,,
\end{equation}
and relate the remaining constants on the right-hand side to state
parameters via 
\begin{equation}
c_1-c_2 = 2(\Delta H)^2 
\quad,\quad c_1+c_2= 2c_1-(c_1-c_2)= 2(A^2-H^2-(\Delta H)^2) 
\end{equation}
using (\ref{c1c2}) and (\ref{c1}).

Of particular interest is the asymmetry of volume fluctuations at
$\lambda\to\infty$ compared with $\lambda\to-\infty$. In the cosmological
model based on the present example, this asymmetry corresponds to the relation
between fluctuations long before and long after the ``bounce'' when
$V(\lambda)$ in (\ref{VJ}) is minimal at $\lambda=\lambda_0$. It is given by
the absolute value of the difference of $c_3$ and $c_4$, which we can obtain
from the sum (\ref{c3plusc4}) using (\ref{Uncert}):
\begin{eqnarray} \label{c3c4}
 (c_3-c_4)^2= (c_3+c_4)^2-4c_3c_4&=& \frac{H^4}{A^4}(c_1-c_2)^2+
 \frac{H^2}{A^2}(c_1^2-c_2^2)- H^2\hbar^2\\
&=& 4H^2\left(\left(1-\frac{H^2}{A^2}\right)(\Delta H)^2-
  \frac{\hbar^2}{4}+ \left(\frac{H^2}{A^2}-1\right)\frac{(\Delta
    H)^4}{A^2}\right)\,.\nonumber
\end{eqnarray}

The solutions of this subsection tell us how moments evolve, starting
from some initial values that belong to an initial state. If the
initial state saturates the uncertainty relation, it will always do
so: the left-hand side of the relation is a Casimir function on the
space of moments. If we combine our solutions with the results of the
previous subsection, we are guaranteed the existence of dynamical
coherent states, for we know that there is a coherent initial state
whose moments must evolve as derived here. The question, addressed in
\cite{BeforeBB,Harmonic}, is then in how far the coherence of the state can
restrict the asymmetry of fluctuations.

\subsection{Beyond coherent states, and cosmic forgetfulness}

At this stage, we have reproduced the asymmetry derived in \cite{Harmonic}
from all three uncertainty relations in a different way, using reality
conditions and only one of the uncertainty relations. Since we already know
that second-order reality conditions reduce the number of uncertainty
relations to just one, this result is not surprising. However, the
rederivation allows a powerful generalization of the asymmetry formula to all
semiclassical states, not just dynamically coherent ones. Our reality
conditions are valid provided only that moments of order higher than second
are subdominant, which is the most general definition of semiclassical
states. The preceding derivation remains intact if we change the equality in
(\ref{Uncert}) to an inequality once we depart from dynamical coherent
states. In this way, the last formula, (\ref{c3c4}), changes to the
inequality
\begin{equation}
 (c_3-c_4)^2\leq
 4H^2\left(\left(1-\frac{H^2}{A^2}\right)(\Delta H)^2-
  \frac{\hbar^2}{4}+ \left(\frac{H^2}{A^2}-1\right)\frac{(\Delta
    H)^4}{A^2}\right)
\end{equation}
valid for all semiclassical states (or, more generally, whenever moments of
order three and higher can be ignored, even if second-order moments are large
compared to expectation values). In particular, the change of volume
fluctuations is bounded by the quantum fluctuation $\Delta H$ of the
Hamiltonian. (In the cosmological model, $H$ corresponds to the momentum of a
free, massless scalar used to parameterize time.) The inequality derived here
generalizes the identity found in \cite{BounceCohStates} for dynamical
coherent states saturating the uncertainty relation. It is also consistent
with the bound derived in \cite{LoopScattering} for a larger class of states.

The question of asymmetric volume fluctuations has been raised in
\cite{BeforeBB} and led to a lively debate in the literature on loop quantum
cosmology. Even though the asymmetry of volume fluctuations is bounded by
energy fluctuations, ensuring that an initial semiclassical state does not
develop too-large volume fluctuations, the change of fluctuations may
nevertheless be significant. Moreover, it depends sensitively on the initial
values \cite{Harmonic}. If the moments of a state had to be known for
long-term state evolution, the sensitivity and potential significant changes
near the turning point of $\langle\hat{V}\rangle$ render a reliable analysis
of the state at very early times practically impossible, implying cosmic
forgetfulness. (As noted, the issue is not as relevant in recent versions of
the scenario, in which one is required to eliminate deterministic evolution
through the ``bounce'' due to signature change at high density.)

The possibility of significant changes in volume fluctuations becomes
clear when one looks at the relative change of relative volume fluctuations,
for dynamical coherent states given by
\begin{equation}
\left|1-\frac{\lim_{\lambda\to-\infty}(\Delta
V)^2(\lambda)}{\lim_{\lambda\to\infty} (\Delta V)^2(\lambda)}\right|=
\frac{|c_3-c_4|}{c_4}\sim \frac{(\Delta H)^2}{(\Delta V)^2}\,.
\end{equation}
If this value is near zero, volume fluctuations are nearly symmetric. However,
the ratio on the right-hand side depends more sensitively on the precise state
and cannot be restricted to be small without further assumptions. Moreover, in
cosmology it is usually a safe assumption that matter behaves more quantum
than geometry; thus, the energy fluctuation should be expected to be
significantly larger than volume fluctuations and the right-hand side is much
larger than one. Large changes of volume fluctuations are not ruled out,
implying, together with the large sensitivity to initial values, that the
early state cannot be reconstructed precisely even if evolution were
deterministic.

\section{Conclusions}

We have presented several constructions and results for moments of states in
quantum systems corresponding to irreducible representations of groups. The
Casimir conditions that select a representation can be dealt with using
methods for effective first-class constrained systems. In contrast to the
general case of first-class constraints, simplifications occur that allowed us
to draw conclusions for arbitrary orders in a semiclassical or moment
expansion. Our effective methods are then particularly useful to derive
representation-independent relationships between moments of group coherent
states. Some of our results on uncertainty relations can be applied more
generally, even when no Casimir condition is imposed.

These methods are useful whenever it is difficult to construct explicit wave
functions for group (or other) coherent states. We have shown this in our
cosmological example, in which the question of cosmic forgetfulness requires
good control on all possible coherent states, not just on those of small
squeezing which are easier to construct as wave functions. When no assumption
on the amount of squeezing is made, even a coherent state does not allow good
control on quantum parameters such as fluctuations compared over long time
intervals. 

As also shown in this example, it is possible to show that coherent states
exist even without knowing their wave functions. Effective equations for
moments then allow one to compute the corresponding quantum
parameters. Results obtained by these effective methods are no less reliable
than those of more cumbersome calculations using explicit wave functions.

\section*{Acknowledgements}

We thank David Brizuela for discussions. This work was supported in part by
NSF grants PHY-0748336 and PHY-1307408.

\begin{appendix}

\section{Ordering relations}

Some of the equations in this paper can be reproduced by making use of the
following relationships:
\begin{eqnarray}
\hat{A}\hat{B}\hat{C} &=& (\hat{A}\hat{B}\hat{C})_{\rm Weyl}-\frac{1}{4}
\left(\{\hat{A},[\hat{C},\hat{B}]\}+ \{\hat{B},[\hat{C},\hat{A}]\}+
  \{\hat{C},[\hat{B},\hat{A}]\}\right)\\
&&+\frac{1}{6} \left([\hat{B},[\hat{C},\hat{A}]]-
  2[\hat{A},[\hat{C},\hat{B}]]\right)\nonumber
\end{eqnarray}
and
\begin{eqnarray}
\hat{A}\hat{B}\hat{C}\hat{D} &=& (\hat{A}\hat{B}\hat{C}\hat{D})_{\rm Weyl}\\
&& -\frac{1}{2}\left((\hat{C}\hat{D}[\hat{B},\hat{A}])_{\rm Weyl}+
  (\hat{B}\hat{D}[\hat{C},\hat{A}])_{\rm Weyl}+
  (\hat{B}\hat{C}[\hat{D},\hat{A}])_{\rm Weyl}\right.\nonumber\\
&&+\left.
  (\hat{A}\hat{C}[\hat{D},\hat{B}])_{\rm Weyl}+
  (\hat{A}\hat{D}[\hat{C},\hat{B}])_{\rm Weyl}+
  (\hat{A}\hat{B}[\hat{D},\hat{C}])_{\rm Weyl}\right)\nonumber\\
&& +\frac{1}{8} \left(\{[\hat{A},\hat{B}],[\hat{C},\hat{D}]\}+
  \{[\hat{A},\hat{C}],[\hat{B},\hat{D}]\}+
  \{[\hat{A},\hat{D}],[\hat{B},\hat{C}]\}\right)\nonumber\\
&&-\frac{1}{12} \left(2\{\hat{C},[[\hat{B},\hat{A}],\hat{D}]\}-
  \{\hat{C},[[\hat{D},\hat{A}],\hat{B}]\}+
  2\{\hat{D},[[\hat{B},\hat{A}],\hat{C}]\}-
  \{\hat{D},[[\hat{C},\hat{A}],\hat{B}]\}\right.\nonumber\\
&&+\left.
  2\{\hat{B},[[\hat{C},\hat{A}],\hat{D}]\}-
  \{\hat{B},[[\hat{D},\hat{A}],\hat{C}]\}+
  2\{\hat{A},[[\hat{C},\hat{B}],\hat{D}]\}-
  \{\hat{A},[[\hat{D},\hat{B}],\hat{C}]\}\right)\nonumber\\
&&-\frac{1}{12} \left([\hat{D},[[\hat{C},\hat{B}],\hat{A}]]+
  [\hat{C},[[\hat{D},\hat{B}],\hat{A}]]+
  [\hat{B},[[\hat{D},\hat{C}],\hat{A}]]-
  2[\hat{A},[[\hat{D},\hat{C}],\hat{B}]]\right)\nonumber
\end{eqnarray}
using (only here) anticommutators $\{\cdot,\cdot\}$ to denote
symmetrization. Practically, these relations can be derived somewhat
systematically by starting with the totally symmetrically ordered
$(\hat{A}\hat{B}\cdots)_{\rm Weyl}$ and applying suitable commutators to
arrive at the ordering $\hat{A}\hat{B}\cdots$, upon which lower-order
relations can be used to symmetrize all remaining terms. (Note that we write
expressions such as $(\hat{C}\hat{D}[\hat{B},\hat{A}])_{\rm Weyl}$ with the
understanding that the commutator $[\hat{B},\hat{A}]$ has been evaluated
before symmetrization is applied.)

The structure of these relations, with terms making use of different
compositions of symmetrization and antisymmetrization, follows more easily
from Young tableaux. However, the standard Young projectors write
antisymmetrizations by summing over permutation groups, rather than by
iterated applications of commutators. This latter form is more useful for our
purposes and would have to be derived from full antisymmetrizations if Young
tableaux were used.

\end{appendix}

%\bibliographystyle{../preprint}
%\bibliography{../Bib/QuantGra}

\end{document}